\documentclass{jfmarxiv}

\usepackage{amsmath,amssymb}
\usepackage{graphicx}
\usepackage{natbib}
\usepackage{xcolor}
\usepackage{color}

\ifCUPmtlplainloaded \else
 \checkfont{eurm10}
 \iffontfound
\IfFileExists{upmath.sty}
  {\typeout{^^JFound AMS Euler Roman fonts on the system,
               using the 'upmath' package.^^J}%
   \usepackage{upmath}}
  {\typeout{^^JFound AMS Euler Roman fonts on the system, but you
               dont seem to have the}%
   \typeout{'upmath' package installed. JFM.cls can take advantage
             of these fonts,^^Jif you use 'upmath' package.^^J}%
   \providecommand\upi{\pi}%
  }
 \else
\providecommand\upi{\pi}%
 \fi
\fi

\ifCUPmtlplainloaded \else
 \checkfont{msam10}
 \iffontfound
\IfFileExists{amssymb.sty}
  {\typeout{^^JFound AMS Symbol fonts on the system, using the
            'amssymb' package.^^J}%
   \usepackage{amssymb}%
   \let\le=\leqslant  
     
  }{}
 \fi
\fi

\ifCUPmtlplainloaded \else
 \IfFileExists{amsbsy.sty}
{\typeout{^^JFound the 'amsbsy' package on the system, using it.^^J}%
 \usepackage{amsbsy}}
{\providecommand\boldsymbol[1]{\mbox{\boldmath $##1$}}}
\fi

\newcommand{\Real}{\mbox{Re}} 
\newcommand{\Reyn}{\text{\textit{Re}}}

\newcommand{\Schm}{\text{\textit{Sc}}}
\newcommand{\Rich}{\text{\textit{Ri}}}
\newcommand{\Stro}{\text{\textit{St}}}
\newcommand{\Gras}{\text{\textit{Gr}}}
\newcommand{\DoTO}{{D/\theta_0}}
\newcommand{\He}{{\text{He}}}         
\newcommand{\NN}{{\text{N}_2}}        

\newcommand{\ie}{i.e.\ }
\providecommand\bnabla{\boldsymbol{\nabla}}
\providecommand\bcdot{\boldsymbol{\cdot}}
\newcommand{\vect}[1]{\boldsymbol{#1}}
\newcommand{\tens}[1]{\mathsfbi{#1}}

\newcommand{\ui}{\mathrm{i}}
\newcommand{\ur}{\mathrm{r}}
\newcommand{\ud}{\mathrm{d}}
\providecommand{\upi}{\upright{\pi}}

\newcommand{\ms}{\kern.10em\relax}
\newcommand{\fr}[2]{\ensuremath{\frac{#1}{#2}}}

\newcommand{\pfr}[2]{\ensuremath{\frac{\partial #1}{\partial #2}}}
\newcommand{\pfi}[2]{\ensuremath{{\partial #1}/{\partial #2}}}
\newcommand{\ub}{\bar{u}}
\newcommand{\vb}{\bar{v}}
\newcommand{\pb}{\bar{p}}
\newcommand{\rhob}{\bar{\rho}}
\newcommand{\uh}{\skew3\hat{u}}
\newcommand{\vh}{\skew3\hat{v}}
\newcommand{\ph}{\skew3\hat{p}}
\newcommand{\rhoh}{\skew3\hat{\rho}}

\newcommand{\qh}{\skew3\hat{q}}
\newcommand{\qt}{\skew3\tilde{q}}

\title%
[Global instability of low-density jets]%
{Global instability of low-density jets}

\author%
[W. Coenen, L. Lesshafft, X. Garnaud and A. Sevilla]%
{W. Coenen$^{1,2}$ %
\thanks{Email address for correspondence: wicoenen@ucsd.edu},\ns
L. Lesshafft$^3$,\ns%
X. Garnaud$^3$
\thanks{Present address: Safran Tech, Rue des Jeunes Bois, 78772, Magny-Les-Hameaux, France}
and A. Sevilla$^1$}

\affiliation{%
$^1$Grupo de Mec\'anica de Fluidos,
Universidad Carlos III de Madrid,\break
Av.~Universidad 30, 28911 Legan\'es (Madrid), Spain\\[\affilskip]
$^2$Department of Mechanical and Aerospace Engineering,
University of California San Diego,\break
9500 Gilman Dr., La Jolla, CA 92093-0411, USA\\[\affilskip]
$^3$Laboratoire d'Hydrodynamique (LadHyX), \'Ecole polytechnique -- CNRS,\break
91128 Palaiseau, France}

\begin{document}

\maketitle


\begin{abstract}
The global stability of laminar axisymmetric low-density jets is investigated in the
low Mach number approximation. The linear modal dynamics is found to be characterised by two
features: a stable \emph{arc branch} of eigenmodes and an isolated eigenmode. Both
features are studied in detail, revealing that, whereas the former is highly sensitive to
numerical domain size and its existence can be linked to spurious feedback from the outflow
boundary, the latter is the physical eigenmode that is responsible for the appearance of
self-sustained oscillations in low-density jets observed in experiments at low Mach numbers. In
contrast to previous local spatio-temporal stability analyses, the present global analysis
permits, for the first time, the determination of the critical conditions for the onset of global
instability, as well the frequency of the associated oscillations, without additional
hypotheses, yielding predictions in fair agreement with previous experimental observations.  It
is shown that under the conditions of those experiments, viscosity variation with
composition, as well as buoyancy, only have a small effect on the onset of instability.
\end{abstract}



\section{Introduction}
\label{sec:introduction}

Submerged jets become globally unstable, achieving a self-sustained oscillatory
state, when their density is sufficiently smaller than that of their
surroundings, as clearly evidenced by many experimental, theoretical and
numerical studies. The phenomenon was first recognised thanks to the pioneering
work of~\citet{MonkewitzSohn}, who demonstrated the existence of a region of
local absolute instability close to the injector of a turbulent heated jet by
means of a quasi-parallel linear stability analysis. The global transition has
been experimentally characterised in detail both for hot jets
\citep{monkewitz1990} and for light jets, where the density difference is due
to the injection of fluid of smaller molecular weight than that of the
ambient~\citep{Kyle1993,Hallberg2006}, and also by means of a number of local
stability analyses which accounted for the origin of the phenomenon with
increasing
detail~\citep{Jendoubi94,Lesshafft2007,Coenen2008,Lesshafft2010,Coenen2012}.
These studies have been complemented by direct numerical
simulations~\citep{Lesshafft2007a} that unambiguously demonstrated the link
between the existence of locally absolutely unstable regions in the near field
of low-density jets and the onset of global self-sustained oscillations.

Recently, several global linear stability analyses of submerged jet
configurations, avoiding the quasi-parallel approximation, have been performed
thanks to the availability of sufficient computational power and the
development of appropriate numerical techniques. \citet{Nichols2011} pioneered
the use of a global approach to study hot and cold compressible jets.
\citet{Garnaud2013PoF,Garnaud2013JFM} considered the case of constant-density
incompressible jets, revealing that global modes are strongly affected by the
domain length and the numerical discretisation, while the frequency response is
robust and explains the origin of the preferred mode in globally stable jets.
In contrast with the case of constant-density jets, the important experiments
of \citet{Hallberg2006} strongly suggest the existence of an isolated eigenmode
responsible for the global transition for sufficiently low values of the
Reynolds number. Isolated eigenmodes in low-density jets have indeed been
detected by \cite{Nichols2010} for supersonic cases, and by \cite{Qadri2014}
for a low Mach number configuration.

The main objective of the present work is to provide a detailed
characterisation of the global stability properties of light $\He/\NN$ laminar
jets in the low Mach number limit, by means of modal and frequency response
analyses. Two key questions addressed are whether there exists an isolated
eigenmode that explains the experimentally observed transition in low-density
jets, and what are the differences between the global stability properties of
constant-density and low-density jets. 

The paper begins with the mathematical formulation in \S 2. In \S 3 we study a
slowly developing globally stable jet, followed by an analysis of a rapidly
spreading helium jet in \S 4. Finally, concluding remarks are given in \S 5.


\section{Formulation}
\label{sec:formulation}

We consider an axisymmetric laminar low-density $\He$/$\NN$ jet, discharging
with a constant flow rate $Q^\ast$ from an injector pipe with radius $R^\ast$
into an ambient of $\NN$. The ratio of the density $\rho_j^\ast$ of the jet and
the ambient density $\rho_\NN^\ast$ is given by $S=W_j^\ast/W_{\NN}^\ast$. Here
$W_j^\ast = [Y_j/W_\He^\ast + (1-Y_j)/W_\NN^\ast]^{-1}$ is the mean molecular
weight of the jet mixture, determined by the initial mass fraction $Y_j$ of
$\He$. In other terms, to obtain a jet with jet-to-ambient density ratio $S$,
an initial mass fraction $Y_j = (S^{-1}-1)/(W_\NN^\ast/W_\He^\ast-1)$ of $\He$
is injected.  The viscosity $\mu_j^\ast$ of the jet can be related to
$\mu_\He^\ast$ and $\mu_\NN^\ast$ through \citet{Hirschfelder1954}'s law
\citep[see][eq.~2.11]{Coenen2012}. Note that in the formulation dimensional
quantities are indicated with an asterisk $^\ast$. The jet exit values
$\rho_j^\ast$, $\mu_j^\ast$ are used as scales for the dimensionless density
$\rho$ and viscosity $\mu$, whereas the jet radius $R$ is used as the
characteristic length scale, yielding the dimensionless cylindrical coordinate
system $(x,r)$.  The velocity field $\vect{u}=(u,v)$ is non-dimensionalised with
the mean velocity $U_m^\ast = 4 Q^\ast / (\upi {R^\ast}^2)$.  All flow
quantities are taken to be independent of the azimuth $\phi$ throughout this
study; only axisymmetric perturbations are considered.

The Reynolds number of the jet is assumed to be large, $\Reyn = \rho_j^\ast
U_m^\ast R^\ast / \mu_j^\ast \gg 1$, resulting in a slender jet flow. The
importance of buoyancy effects can be estimated through the Richardson number
$\Rich = (\rho_\NN^\ast - \rho_j^\ast) g^\ast R^\ast / (\rho_j^\ast
{U_m^\ast}^2)$. For comparison with experiments, it is useful to write this as
$\Rich = \Gras/\Reyn^2$, where
$\Gras=\rho_j^\ast(\rho_\NN^\ast-\rho_j^\ast)g^\ast{R^\ast}^3/{\mu_j^\ast}^2
=(1/S-1)g^\ast{R^\ast}^3/\nu_j^\ast$ is a Grashof number. The latter only
depends on the injector radius and the properties of the gas mixture, which
usually do not vary within the same experimental campaign. For simplicity, we
will neglect buoyancy effects, \ie we will assume $\Rich \ll 1$, except for the
comparison with experiments in \S{}\ref{sec:resultscompexp}.  Furthermore, for
the description of the resulting jet flow it is assumed that the characteristic
jet velocity $U_m^\ast$ is much smaller than the ambient speed of sound, so
that the simplifications associated with the low Mach number approximation
\citep{Williams1985,Nichols2007} can be applied. This implies that the density
variations in the jet are only due to variations in molecular weight, and are
not related to pressure variations. It also means that if the jet discharges
with the same temperature as the ambient, the flow will remain isothermal
everywhere, and the energy equation is not needed in the description.
Furthermore, in the low Mach number limit, the viscous stress term that is
proportional to the second coefficient of viscosity can be incorporated in the
definition of the variable $p$. This $p$ represents the pressure difference
from the unperturbed ambient distribution, scaled with the characteristic
dynamic pressure $\rho_j^\ast {U_m^\ast}^2$. The jet is then effectively
described by the continuity, momentum conservation and species conservation
equations,
\begin{align}
\pfr{\rho}{t} + \bnabla\bcdot(\rho\vect{u}) & = 0 \ms,
\label{eq:lmnscont}\\
\rho \left( \pfr{\vect{u}}{t} + \vect{u}\bcdot\bnabla\vect{u} \right) & =
- \bnabla p
    + \fr{1}{\Reyn} \bnabla^2\vect{u}
    + \Rich \frac{1-S\rho}{1-S} \vect{e}_x \ms,
\label{eq:lmnsmom}\\
\rho \left( \pfr{Y}{t} + \vect{u}\bcdot\bnabla Y \right) & =
\fr{1}{\Reyn\ms\Schm} \bnabla\bcdot(\rho\bnabla Y) \ms,
\label{eq:lmnsspec}
\end{align}
together with the relation between the mass fraction $Y$ of $\He$ and the
density $\rho$ of the jet,
\begin{equation}
Y = Y_j (1/\rho - S)/(1-S) \ms.
\label{eq:massfracdens}
\end{equation}
In \eqref{eq:lmnsspec} the Schmidt number $\Schm = \mu_j^\ast/(\rho_j^\ast
\cal{D}^\ast)$ is based on the values of the viscosity and density at the jet
exit. For example, the two density ratios used in the present work, $S=0.143$
and $S=0.5$, correspond to Schmidt numbers $\Schm = 1.69$ and $\Schm=0.49$,
respectively. For simplicity, the variation of the viscosity with the composition
is not taken into account in the conservation
equations~\eqref{eq:lmnscont}-\eqref{eq:lmnsspec}, and is thus neglected in the
results, except for the comparison with experiments of
\S{}\ref{sec:resultscompexp}, where its influence is studied separately. To the
latter aim, the viscous term in \eqref{eq:lmnsmom} is written as
$\Reyn^{-1} \bnabla \bcdot [\mu (\bnabla \vect{u} + \bnabla \vect{u}^T)]$ and Hirschfelder's
law \citep[see][eq.~2.11]{Coenen2012} is used to relate the dimensionless viscosity
$\mu = \mu^\ast/\mu_j^\ast$ to the mass fraction $Y$.


\subsection{Base flow}
\label{sec:formulationbaseflow}

\begin{figure}
\centering
\includegraphics[width=\textwidth]{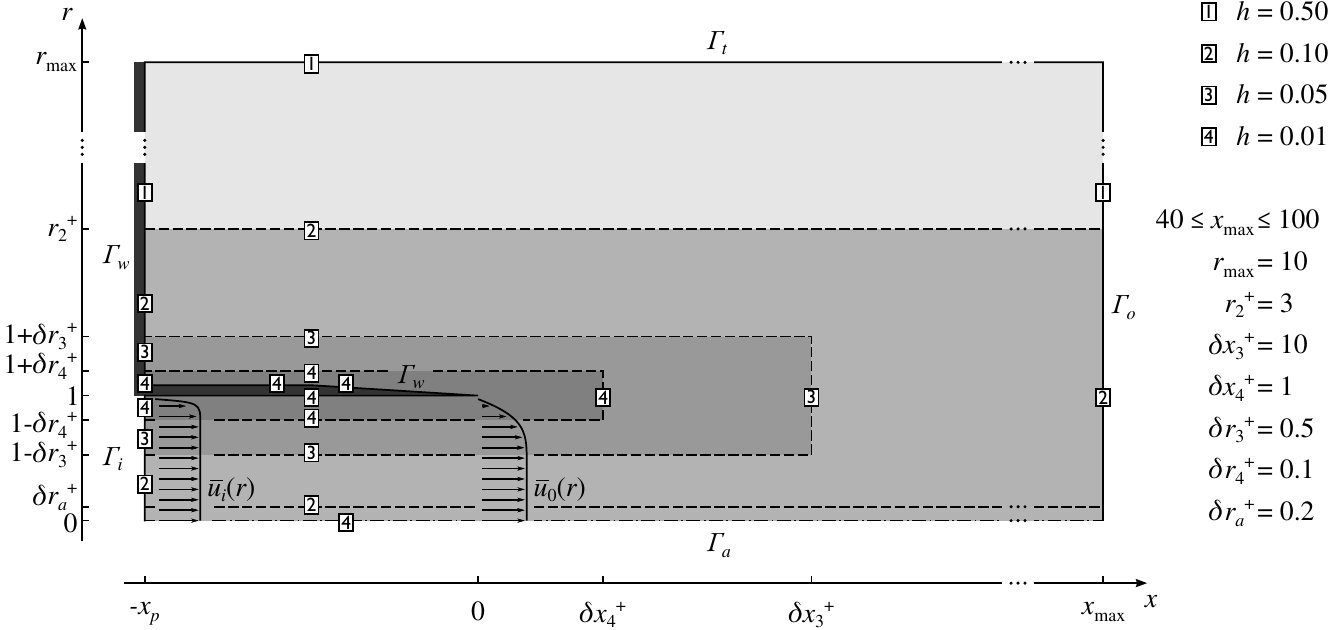}
\caption{%
Schematic representation of the numerical domain. The boxed numbers on the
boundaries indicate the level of refinement through the distance $h$ between
discretisation points. The areas with different grid resolutions that are
obtained in this manner are indicated by the grey shading.}
\label{fig:domain}
\end{figure}

\begin{figure}
\centering
\includegraphics[width=\textwidth]{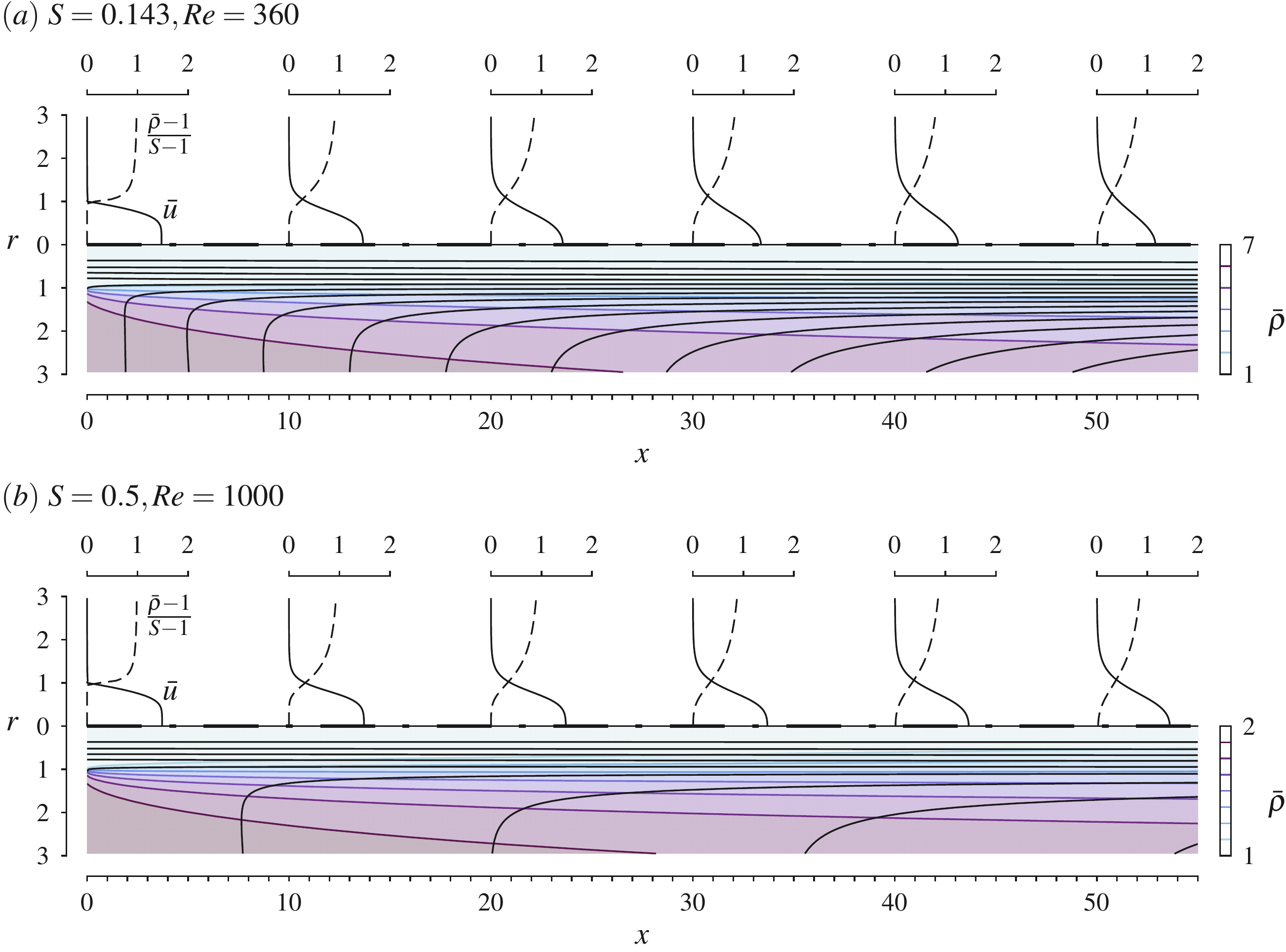}
\caption{%
Radial profiles of axial velocity $\ub$ and rescaled density $(\rhob-1)/(S-1)$
at $x=0, 10, 20, 30, 40, 50$, together with streamlines and contours of the
density field $\rhob$, for $(a)$ $S=0.143$, $\Reyn=360$, $\DoTO=24.3$, and $(b)$
$S=0.5$, $\Reyn=1000$, $\DoTO=24.3$.}
\label{fig:baseflow}
\end{figure}

As a base state for the linear stability analysis, a steady solution
$(\ub,\vb,\pb,\rhob)$ of the low Mach number Navier-Stokes
equations~\eqref{eq:lmnscont}-\eqref{eq:lmnsspec} is employed. The flow domain
under consideration is depicted in figure~\ref{fig:domain}. In order to mimic
experimental conditions, a short injector pipe of length $x_p$ is included,
bounded by a wall $\Gamma_w$ of thickness 0.02 that ends on a 5.7$^\circ$ knife
edge. At the pipe inlet $\Gamma_i$ a velocity profile $\ub_i(r)$ is imposed,
taken from a collection of profiles that is obtained by solving the laminar
boundary-layer equations in a circular pipe \cite[see][\S{}2.1.1]{Coenen2012}.
Because the flow in the short injector pipe of the domain will further develop,
the velocity profile $\ub_i(r)$ at the inlet must be carefully chosen to obtain
a certain desired velocity profile $\ub_0(r)$ at the jet exit plane. We will
characterise the latter profile through the inverse of its dimensionless
momentum thickness $\DoTO$, defined by
\begin{equation}
    \theta_0 =
    \int_0^\infty \frac{\ub_0(r)}{\ub_0(0)}
    \left[1-\frac{\ub_0(r)}{\ub_0(0)}\right] \ms \ud r,
\end{equation}
and scaled with the jet diameter $D=2R$ for consistency with
\citet{Hallberg2006}.  For the computations of the results of the present work
a pipe length $x_p=3$ was used, with the exception of the transition points
with $\DoTO > 30$ of figure~\ref{fig:compexperiments}, for which a pipe length
$x_p=1.5$ was used.  It was verified that larger values of $x_p$ did not affect
the results.  The remaining boundaries of the flow domain are the axis
$\Gamma_a$, the lateral boundary $\Gamma_t$ and the downstream outlet boundary
$\Gamma_o$, as labelled in figure~\ref{fig:domain}.  Stress-free boundary
conditions are imposed on the latter two.  Because the jet entrains fluid
through the lateral boundary, the density at that boundary must be fixed to
$1/S$ to obtain the desired jet-to-ambient density ratio. The radial extent of
the domain was set to $r_\text{max}=10$, and it was checked that larger values
did not change the results. The downstream extent $x_\text{max}$ does not
influence the computation of the base flow, but it does affect various results
of the stability analysis, as will be explained in detail in
\S{}\ref{sec:Re1000} and \S{}\ref{sec:Re360}; values in the range $40 \le
x_\text{max} \le 100$ were used. To summarise, the boundary conditions for the
base flow are
\begin{align}
    \ub - \ub_i = \vb = \rhob - 1 &= 0 \quad\text{on}\quad \Gamma_i, \\
    \ub = \vb = \vect{n}\bcdot\bnabla\rhob &= 0 \quad\text{on}\quad \Gamma_w,\\
    \rhob - \frac{1}{S} =
    -\pb\vect{n} +\frac{1}{\Reyn} \vect{n}\bcdot\bnabla\binom{\ub}{\vb} &= 0
    \quad\text{on}\quad \Gamma_t \\
    -\pb\vect{n} +\frac{1}{\Reyn} \vect{n}\bcdot\bnabla\binom{\ub}{\vb} &= 0
    \quad\text{on}\quad \Gamma_o, \\
    \vb = \partial_r \ub = \partial_r \rhob & = 0 \quad\text{on}\quad \Gamma_a,
\end{align}
where $\vect{n}$ is the outward normal vector on the boundary.

The governing equations are discretised using Taylor-Hood elements, quadratic (P2)
for the density and velocity, and linear (P1) for the pressure, to satisfy the
Ladyzenskaja-Babu\v{s}ka-Brezzi condition. The
refinement of the unstructured mesh is controlled through the distance $h$
between discretisation points on the boundaries of the domain and on auxiliary
lines, as indicated in figure~\ref{fig:domain}. It was checked that the results
were converged with respect to further mesh refinements. The steady base flow is
computed using a Newton-Raphson method and the FreeFem++ software
\citep{Hecht2012}. Figure~\ref{fig:baseflow} shows the resulting base flow for
the two cases that will be studied in detail in \S{}\ref{sec:Re1000} and
\S{}\ref{sec:Re360}: $S=0.143$, $\Reyn=360$, $\DoTO=24.3$, and $S=0.5$,
$\Reyn=1000$, $\DoTO=24.3$.


\subsection{Direct eigenmodes}
\label{sec:formulationdirect}

All experimental and numerical evidence indicate that the global instability
in light jets gives rise to axisymmetric flow oscillations, and the present
analysis therefore is restricted to axisymmetric disturbances. All flow
quantities are independent of the azimuthal coordinate, and the azimuthal
velocity is always zero. Small unsteady axisymmetric perturbations are
introduced into the steady base flow as $(u,v,\rho,p) = (\ub,\vb,\rhob,\pb)+
\varepsilon (u',v',\rho',p')$.  The evolution of these perturbations (primed
quantities) is then governed to leading order by the equations
(\ref{eq:lmnscont}--\ref{eq:massfracdens}), linearised around the base flow,
\begin{align}
\bnabla \bcdot \vect{u}' & =
- \frac{1}{\Reyn\ms\Schm} \bnabla \bcdot \left( \frac{1}{\rhob} \bnabla \rho'
- \frac{\rho'}{\rhob^2} \bnabla\rhob \right) \,,
\label{eq:stabeq1} \\
\pfr{\vect{u}'}{t}
+ \vect{\ub} \bcdot \bnabla \vect{ u}'
+ \vect{ u}' \bcdot \bnabla \vect{\ub}
+ \frac{\rho'}{\rhob} \vect{\ub} \bcdot \bnabla \vect{\ub} & =
- \frac{1}{\rhob} \bnabla p'
+ \frac{1}{\Reyn} \frac{1}{\rhob} \nabla^2 \vect{u}'
- \Rich \frac{S}{1-S} \frac{\rho'}{\rhob} \vect{e}_x \,,
\label{eq:stabeq2} \\
\pfr{\rho'}{t}
+ \vect{\ub} \bcdot \bnabla \rho'
+ \vect{ u}' \bcdot \bnabla \rhob
- \frac{\rho'}{\rhob} \vect{\ub} \bcdot \bnabla \rhob & =
- \frac{1}{\Reyn\ms\Schm} \ms \rhob \ms
\bnabla \bcdot \left( \frac{1}{\rhob} \bnabla \rho'
    - \frac{\rho'}{\rhob} \bnabla \rhob \right) \,.
\label{eq:stabeq3}
\end{align}
Assuming temporal normal-mode solutions
\begin{equation}
(u',v',\rho',p') = (\uh,\vh,\rhoh,\ph)\ e^{-\ui\omega t} \, ,
\label{eq:normalmodes}
\end{equation}
the linearised equations can be written in the form of a generalised eigenvalue
problem
\begin{equation}
-\ui\omega \mathcal{B} \vect{\qh} = \mathcal{L} \vect{\qh} \, ,
\label{eq:EVP}
\end{equation}
where $\vect{\qh}(r,x) = [\uh(r,x),\vh(r,x),\rhoh(r,x),\ph(r,x)]^T$ is the
vector-valued eigenfunction that contains all perturbation quantities.
For what follows, let $\vect{\qt}$, $\tens{L}$
and $\tens{B}$ be understood to be the eigenvector and the matrices of the
\emph{discretised} eigenvalue problem, $-\ui\omega \tens{B} \vect{\qt} =
\tens{L} \vect{\qt}$, that is to be solved numerically.

The following boundary conditions are imposed for the perturbation variables
(boundary labels as given in figure \ref{fig:domain}):
\begin{align}
    \uh = \vh &= 0 \quad\text{on}\quad \Gamma_i, \Gamma_w \, , \\
    -\ph\vect{n} + \frac{1}{\Reyn}\vect{n}\bcdot\bnabla\binom{\uh}{\vh} &= 0
        \quad\text{on}\quad \Gamma_t, \Gamma_o \, , \\
    \vh = \partial_r \uh & = 0 \quad\text{on}\quad \Gamma_a \, .
\end{align}

The discrete system matrices  $\tens{L}$ and $\tens{B}$ are constructed with a
finite element formalism in FreeFEM++, analogous to the incompressible
computations by \cite{Garnaud2013PoF}, using P2 elements for $\rhoh$, $\uh$,
$\vh$ and P1 elements for $\ph$. These matrices are then exported to Matlab for
the solution of the eigenvalue problem, by use of the ARPACK library, and for
all further post-processing. The eigenvalue computation involves an LU
decomposition for inversion of the shifted system.


\subsection{Adjoint eigenmodes}
\label{sec:formulationadjoint}

The physical discussion of eigenmode dynamics in \S{}\ref{sec:Re360} will be
based on the structural sensitivity formalism proposed by \cite{GL07}. Such an
analysis requires the computation of the adjoint discrete eigenvector
$\vect{\qt}^\dag$ associated with a given eigenvalue $\omega$ of the direct
problem (\ref{eq:EVP}). The form of the adjoint eigenvalue problem, of which
$\vect{\qt}^\dag$ is a solution, depends on the definition of an inner product.
Let the inner product between two perturbation states $\vect{\qh}_1$ and
$\vect{\qh}_2$ be defined as the standard spatial integral in cylindrical
coordinates,
\begin{equation}
\langle \vect{\qh}_1, \vect{\qh}_2 \rangle
    = \int_\Omega (
            \uh_1^* \uh_2 +
            \vh_1^* \vh_2 +
            \rhoh_1^* \rhoh_2 +
            \ph_1^* \ph_2) \ms r \ms \ud r \ms \ud x
    = \vect{\qt}_1^H \tens{Q} \vect{\qt}_2 \, .
\end{equation}
Again, the symbols $\vect{\qh}_{1,2}$ are meant to represent the continuous
spatial distribution of perturbations, whereas $\vect{\qt}_{1,2}$ represent the
discretised form, containing all $N$ degrees of freedom of the discrete
problem. The $N\times N$ matrix $\tens{Q}$ contains the metric coefficients for
a given spatial discretisation, reflecting the area size of the individual mesh
elements as well as the weight factor $r$ from the integral. It is a diagonal,
positive definite Hermitian matrix.

With this definition, the discrete adjoint eigenvalue problem is found to be
\begin{equation}
\ui\omega^\dag\tens{Q}^{-1}\tens{B}^H\tens{Q} \vect{\qt}^\dag
    = \tens{Q}^{-1}\tens{L}^H\tens{Q} \vect{\qt}^\dag \, .
\end{equation}
Each adjoint eigenvalue $\omega^\dag$ is the complex conjugate of an associated
direct eigenvalue $\omega$.

For the presentation of results in \S{}\ref{sec:Re1000} and
\S{}\ref{sec:Re360}, direct eigenvectors
are always normalised such that
\begin{equation}
\|\vect{\qt}\|^2 = \vect{\qt}^H \tens{Q} \vect{\qt} = 1 \, ,
\end{equation}
whereupon the associated adjoint eigenvectors are normalised according to
\begin{equation}
\vect{\qt}^{\dag H} \tens{Q} \tens{B} \vect{\qt} = 1 \, .
\end{equation}


\subsection{Eigenvalue sensitivity}
\label{sec:formulationsens}

The sensitivity of an eigenvalue measures how much the eigenvalue is affected by variations of
the associated operator.  According to the procedure proposed by \cite{GL07}, a spatial map of
the sensitivity of $\omega$ with respect to `internal feedback' interactions can be obtained by
measuring the local overlap between the direct and adjoint eigenfunctions. The idea is to
introduce small variations into the system matrix $\tens{L}$ that modify the coupling between
perturbation variables at a given point in space. Several choices are possible to estimate the
effect of such modifications in the \emph{local structure} of the operator on the eigenvalue. We
adopt here the original formulation chosen by \cite{GL07}, which provides an upper-bound
estimation of the eigenvalue drift due to modified velocity-velocity coupling. The
formulation is slightly altered to include the density. It is convenient to first define the $3
\times N$ matrix $\tens{U}_{x_i}$ that extracts from a vector $\vect{\qt}$ the
velocity components and density $[\uh(\vect{x}_i),\vh(\vect{x}_i), \rhoh(\vect{x}_i)]^T =
\tens{U}_{x_i}\vect{\qt}$ at a given discretisation point $\vect{x}_i$. Following the
derivation in \cite{GL07}, the structural sensitivity in the present context is then
characterised by the scalar quantity
\begin{equation}
\lambda(\vect{x}_i)
    = \|\tens{U}_{x_i} \tens{Q} \vect{\qt}^\dag\| \,
      \|\tens{U}_{x_i}\vect{\qt}\| \, .
\label{eqn:structsens}
\end{equation}
\cite{GL07} argue that flow regions with a large value of $\lambda$ influence strongly the
eigenvalue selection, and thus represent the `core' or `wavemaker' of the eigenmode. Additional
information can be obtained by analysing the components of the structural sensitivity tensor
$\tens{S}(\vect{x}_i) = \tens{U}_{x_i} \tens{Q} \vect{\qt}^\dag (\tens{U}_{x_i}\vect{\qt})^H$,
which represent how changes in the feedback from axial velocity, radial velocity, and density
into the axial momentum, radial momentum, and species conservation equation can affect the
eigenvalue~\citep[see, for example,][]{Qadri2015}. Note that the Frobenius norm of $\tens{S}$ is
equal to the structural sensitivity $\lambda$.  It is noted that the choice to only include the
density and velocity in \eqref{eqn:structsens} is rather arbitrary. The quantity $\lambda$ could
just as well be based on momentum, vorticity or combinations thereof, if such a choice appeared
physically more sensible.

\citet{Marquet2008} developed the theoretical framework to assess the
sensitivity of a global eigenmode to arbitrary (not necessarily solution of the
Navier--Stokes equations) modifications of the base flow.
In the present work we will use this concept to study how modifications in the base flow
velocity $\vect{\ub} = (\ub, \vb)$ affect the growth rate $\omega_\text{i}$:
\begin{equation}
    \bnabla_{\vect{\ub}} \, \omega_\text{i}
        = \left(\bnabla_{\ub} \, \omega_\text{i}, \bnabla_{\vb} \, \omega_\text{i}\right)
        = \text{Re} \left[
        -\bnabla (\vect{\uh})^H \bcdot \vect{\uh}^\dag
        + \bnabla \vect{\uh}^\dag \bcdot \vect{\uh}^\ast
    \right],
\end{equation}
where $\vect{\uh}$ and $\vect{\uh}^\dag$ contain the velocity components of
the direct and adjoint eigenmodes.


\subsection{Pseudospectrum}
\label{sec:formulationpseudospec}

As \cite{TrefethenBook} note in their preface, `eigenvalues might be
meaningful in theory, but they [can] not always be trusted on a computer'.
This remark is highly pertinent for the present study: the linearised
Navier--Stokes operator is known to be non-normal \citep[see the review
by][]{chomaz2005}, and this property has important implications for the
sensitivity of eigenmodes with respect to details of the discretisation and to
finite precision arithmetics. In the study of non-normal dynamics, the
\emph{pseudospectrum} provides a very valuable basis for physical discussion.

According to \cite{TrefethenBook}, the $\epsilon$-pseudospectrum can be defined
in at least three equivalent ways. For the purpose of the present study, we
will adopt the definition that a given complex frequency $\omega$ is an
$\epsilon$-pseudo\-eigenvalue of the linear flow equations if
\begin{equation}
\| (\ui \omega \tens{B} + \tens{L})^{-1} \| = \epsilon^{-1} \, .
\end{equation}
The operator $(\ui \omega \tens{B} + \tens{L})^{-1}$ is the \emph{resolvent} of
$\tens{L}$, and its spectral norm is given by its largest singular value
$\sigma$. In physical terms, the largest singular value represents the
\emph{optimal gain} that can be achieved when forcing the system at frequency
$\omega$. We obtain $\epsilon^{-1}=\sigma$ as the leading singular value in the
same way as \cite{Garnaud2013JFM}.



\section{Analysis of a slowly developing stable jet: the arc branch}
\label{sec:Re1000}

The first case to be investigated is a jet of Reynolds number $Re=1000$ and
density ratio $S=0.5$. A shear layer  thickness given by $\DoTO=24.3$ is
measured at the nozzle exit. While $Re=1000$ may seem to be a low value for a
jet, in laminar conditions it yields a very slow viscous spreading,
as can be seen in figure \ref{fig:baseflow}$(b)$.

\begin{figure}
\centering
\includegraphics[scale=0.5]{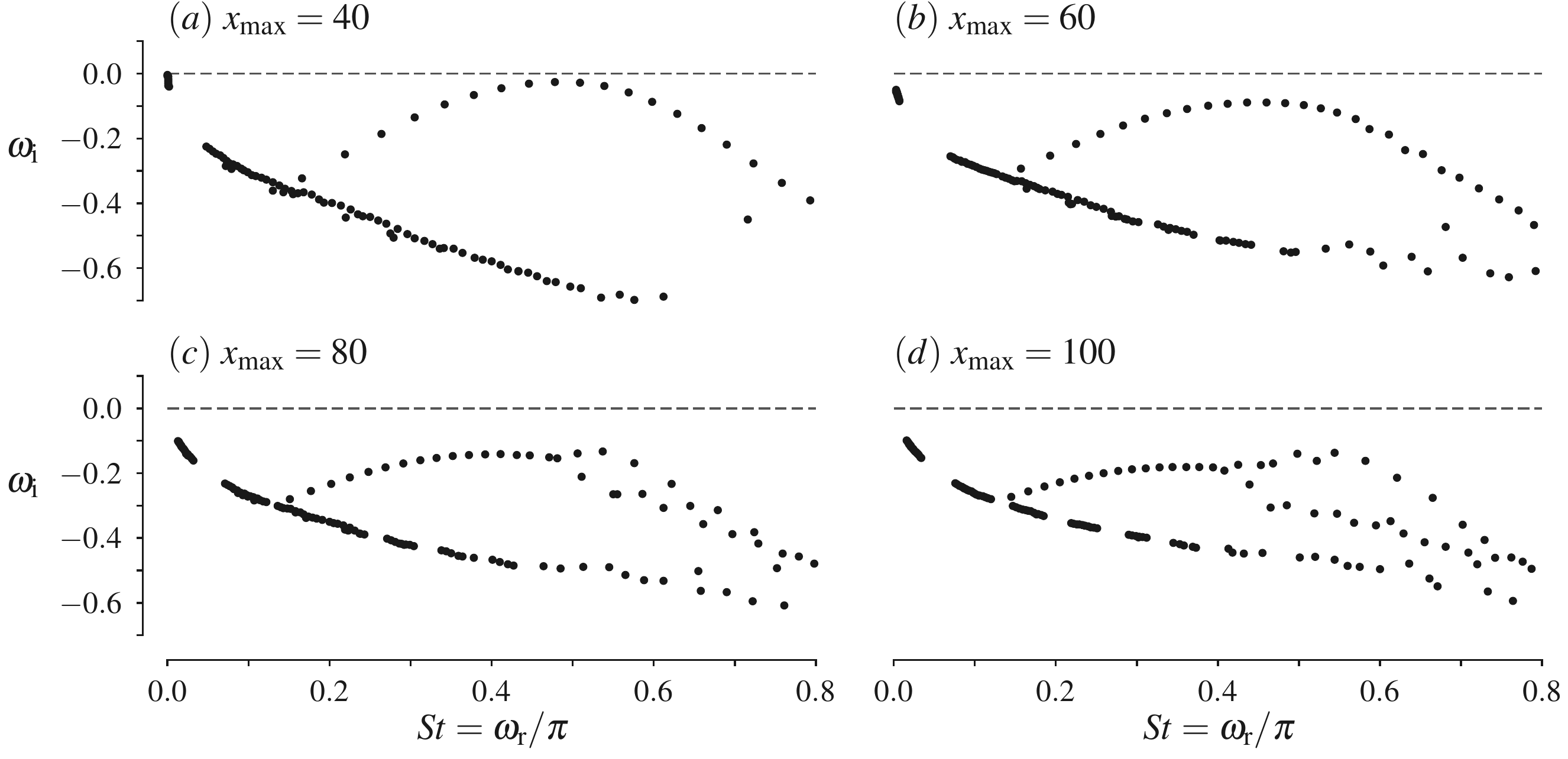}
\caption{%
Eigenvalue spectra of a jet at $Re=1000$, $S=0.5$ and
$\DoTO=24.3$, obtained in numerical domains of different length.}
\label{fig:Re1000spectrum}
\end{figure}

The eigenvalue spectrum for this setting is shown in figure
\ref{fig:Re1000spectrum}, where different panels contain results obtained with
different numerical box lengths $x_\text{max}$ between 40 and 100. The radial
extent in all cases is $r_\text{max}=10$.  The dominant feature of the spectrum
is an upper arching branch of eigenvalues, named the \emph{arc branch} in the
following. Eigenvalues are distributed along it with an even spacing in the
real frequency. All eigenvalues are confined to the stable half-plane of
$\omega$, but in the case of the shortest box length, $x_\text{max}=40$, the
arc branch nearly crosses into the unstable domain.  Clearly, convergence with
respect to the box length $x_\text{max}$ is not achieved. This is consistent
with the analysis of constant-density jets by \cite{Garnaud2013PoF}, who argue
that a further increase of $x_\text{max}$ is not guaranteed to ever lead to
convergence. It seems unreasonable anyway to assume that the linear dynamics
more than 100 radii downstream of the nozzle, in a hypothetical steady flow,
should have any physical relevance.

A lower branch of densely packed eigenvalues is also observed. These modes have
been discussed by \cite{Garnaud2013PoF}, and they will not be investigated in
more depth in this paper.

\begin{figure}
\centering
\includegraphics[scale=0.5]{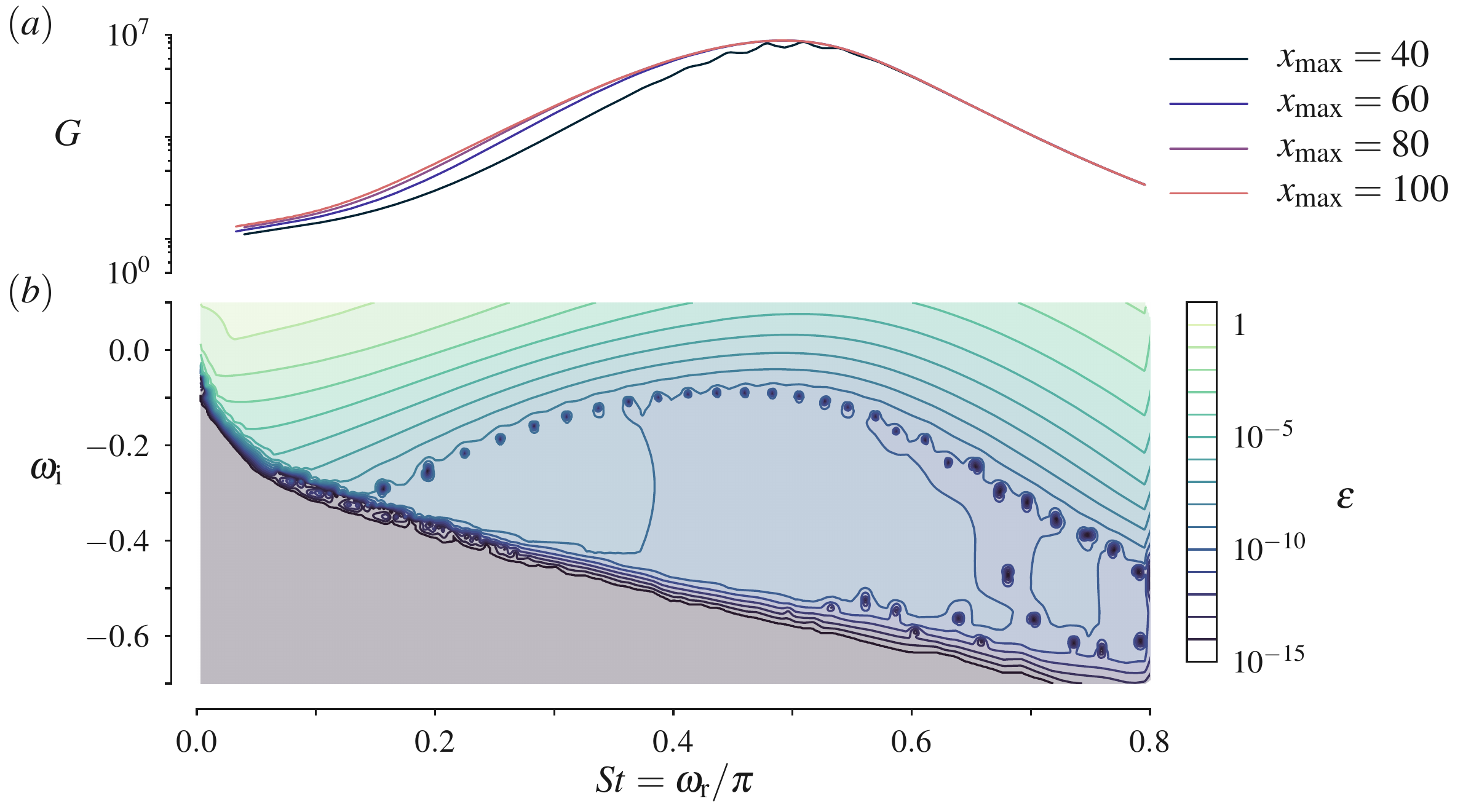}
\caption{%
Forcing response of a jet at $Re=1000$, $S=0.5$ and $\DoTO=24.3$. $(a)$
Optimal gain as a function of real forcing frequency. Different colours denote
results obtained in numerical domains of different lengths, as indicated in the
legend. $(b)$ Pseudospectrum, represented as the inverse of the optimal gain for
complex forcing frequencies, $\epsilon=\sigma^{-1}(\omega)$. The case
$x_\text{max}=60$ is represented.}
\label{fig:Re1000pseudo}
\end{figure}

If the entire spectrum of a flow is stable, its dynamics in a noisy environment
is determined by the response to external forcing. The linear frequency
response of the present jet is represented in figure
\ref{fig:Re1000pseudo}$(a)$ by the optimal energy gain as a function of real
forcing frequency (see \S{}\ref{sec:formulationpseudospec}), as calculated in
numerical domains of different length.  The gain curve is clearly affected by
domain truncation in the case of the shortest box length, $x_\text{max}=40$,
but all curves obtained with larger domains are seen to be in
close agreement.  The maximum gain is reached at $\omega_\ur=1.55$,
corresponding to a value of the Strouhal number based on the jet diameter
$\Stro=\omega_\ur/\pi=0.49$.  It is remarkable that the forced (exogenous)
dynamics is well converged with respect to the box length, while the unforced
(endogenous) dynamics is not. Note also that the maximum gain,
$\mathcal{O}(10^7)$, is very large compared to that of the constant-density
setting, $\mathcal{O}(10^2)$, investigated by \citet{Garnaud2013JFM}, for the
same value of the Reynolds number.  The main difference between the two
configurations lies in the choice of the base state, which in the case of
\citet{Garnaud2013JFM} was a model mean flow with constant density.

The full pseudospectrum of the slowly developing jet is shown in figure
\ref{fig:Re1000pseudo}$(b)$. Two observations are pointed out: firstly, the arc
branch is seen to align approximately with a pseudospectrum contour (here with
a value of approximately $10^{-8}$); secondly, the pseudospectrum variations below the
arc branch are markedly different from those above the branch. Below the arc
branch, the pseudospectrum indeed is nearly constant, compared to the strong
variations in the upper part of the domain.

\begin{figure}
\centering
\includegraphics[width=\textwidth]{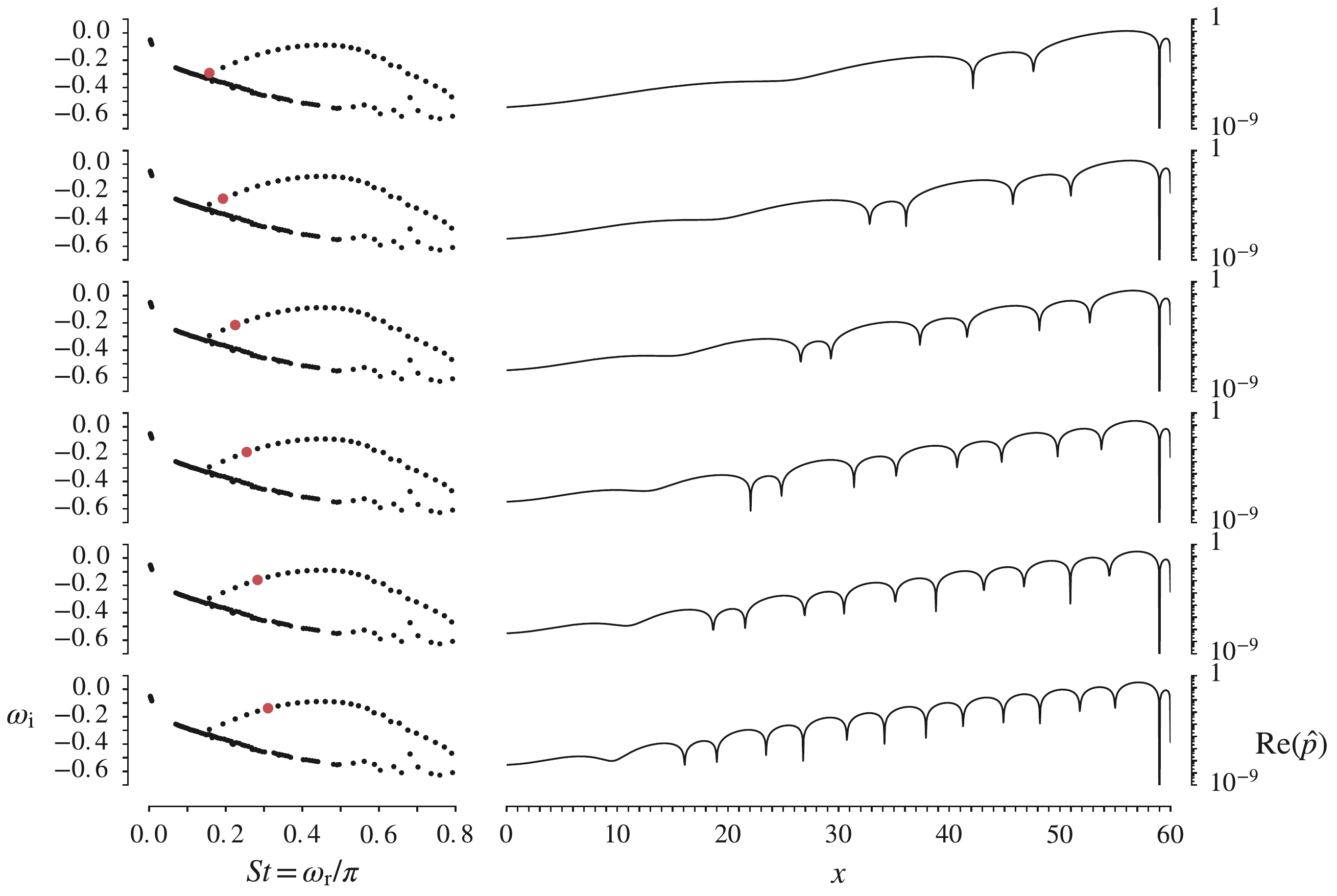}
\caption{%
Pressure eigenfunction along $r=1$ for several modes of the arc branch for
$S=0.5$, $\Reyn=1000$ and $\DoTO = 24.3$, with a numerical domain length
$x_\text{max}=60$.}
\label{fig:boxresonance}
\end{figure}

For a physical interpretation, eigenfunctions of the first six arc branch modes
are represented in figure \ref{fig:boxresonance} by their pressure component
along the nozzle lip line, $r=1$, as a function of $x$. Absolute real values
$|\Real(\hat{p})|$ are plotted in logarithmic scale, and the phase is adjusted
such that $\Real(\hat{p})$ is zero at $x=59$ in all cases. The eigenfunctions
take the form of wavepackets; their particularity lies in the fact that each
mode fits an integer number of wavelengths inside the numerical domain.  Only
the first six eigenfunctions are shown, but the same characteristic applies to
all arc branch modes. The number of wavelengths increases steadily as one moves
along the arc branch from low to higher real frequency values. This observation
suggests that the arc branch is composed of \emph{box modes}, similar to
resonance modes in a pipe of finite length, a conjecture that deserves future investigation.


\section{Analysis of a rapidly spreading pure helium jet}
\label{sec:Re360}

\begin{figure}
\centering
\includegraphics[scale=0.5]{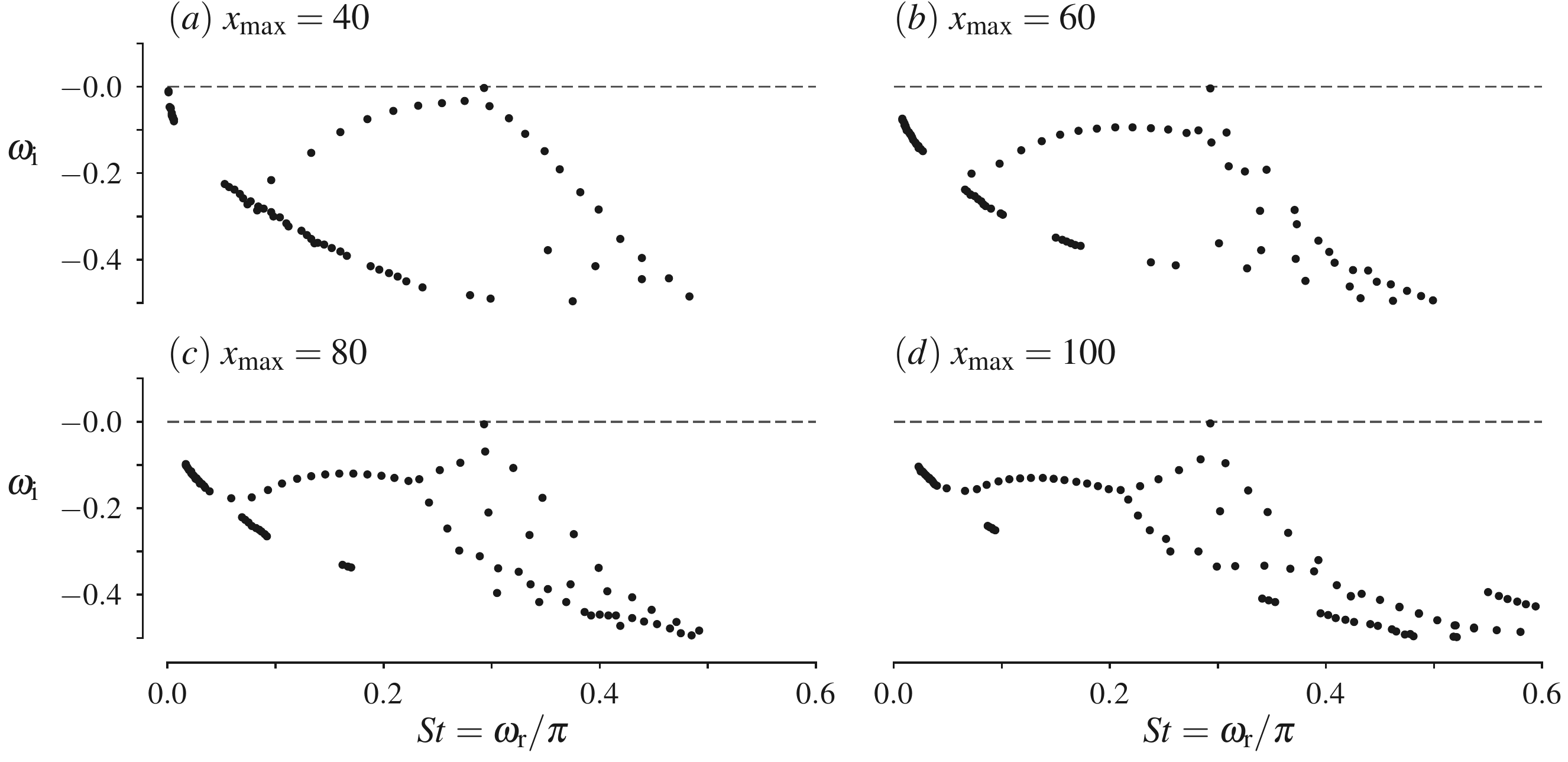}
\caption{%
Eigenvalue spectra of a jet at $Re=360$, $S=0.143$ and $\DoTO=24.3$,
obtained in numerical domains of different length.}
\label{fig:Re360spectrum}
\end{figure}

A jet with parameters $S=0.143$ (pure helium), $\DoTO=24.3$ and $\Reyn=360$ is
considered next. The strong density contrast is certain to result in absolute
instability close to the nozzle \citep{Coenen2012}, and the low value of the
Reynolds number leads to a fast viscous spreading of the jet base flow, as seen
in figures~\ref{fig:baseflow}$(a)$.

The eigenvalue spectrum for this setting is shown in figure
\ref{fig:Re360spectrum}, where different panels again contain results obtained
with different numerical box lengths $x_\text{max}$ between 40 and 100. The
radial extent still is maintained at $r_\text{max}=10$. One single eigenvalue
$\omega=0.9197-0.0042\ui$, very near marginal instability, is identically
obtained (within $|\Delta\omega|=0.0015$) independently of $x_\text{max}$. This
eigenmode, indeed the only one that seems to be converged with respect to box
size, will be denoted here as the \emph{isolated mode}.

An arc branch can be discerned, similar to the one described in
\S{}\ref{sec:Re1000}. In the spectrum of the shortest numerical box,
figure~\ref{fig:Re360spectrum}$(a)$, the eigenvalues
along this branch are still evenly distributed. With
larger box lengths, this regularity persists at low frequencies, but breaks
down in the vicinity of the isolated mode. The pattern observed here resembles
the ``zipper phenomenon'', described by \cite{Heaton2009} and
\cite{Nichols2011b}.


\subsection{The isolated mode}
\label{sec:isolatedmode}

\begin{figure}
\centering
\includegraphics[width=\textwidth]{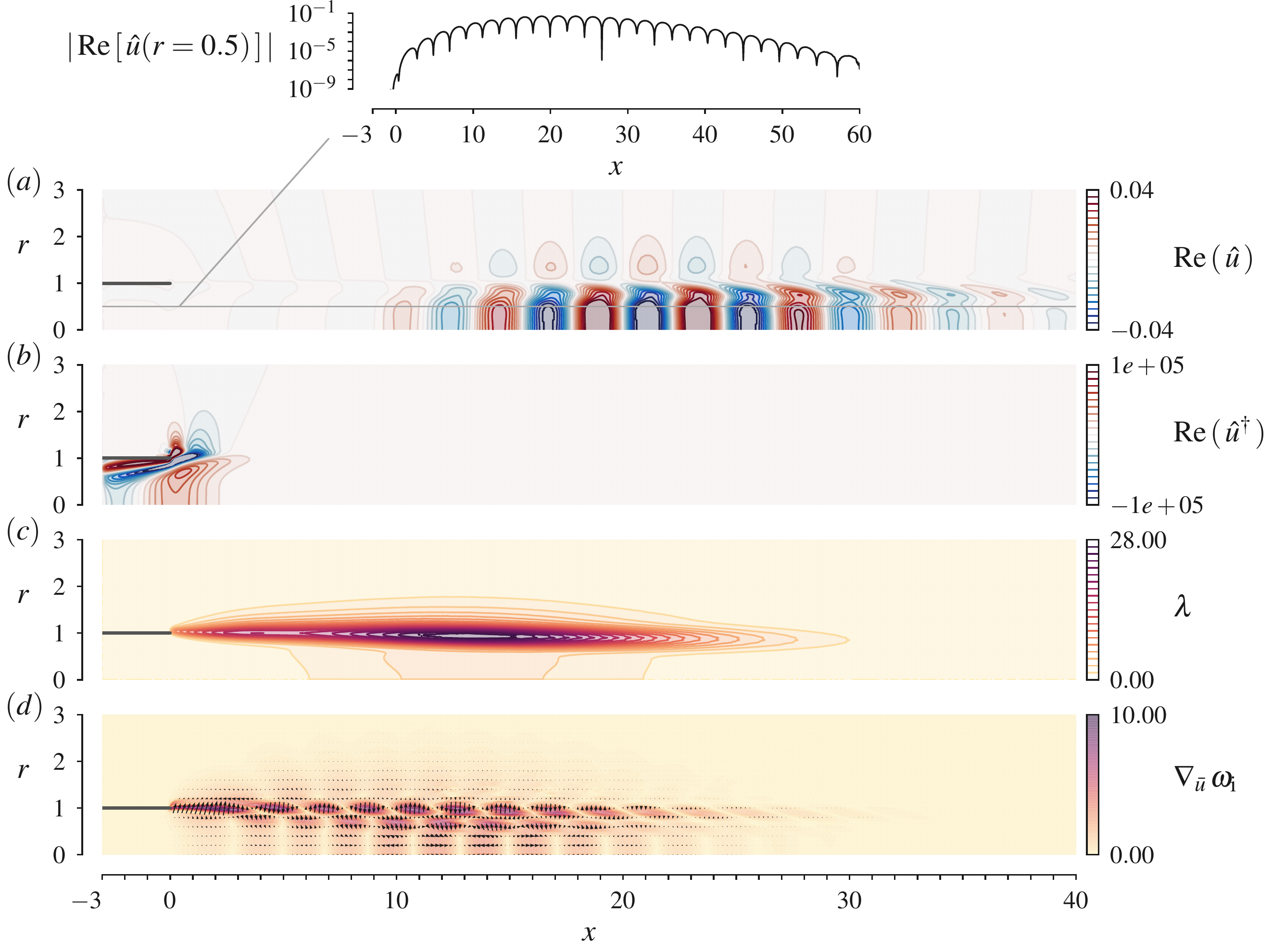}
\caption{%
$(a)$ Spatial structure of the eigenfunction of axial velocity $\uh$
corresponding to the most unstable eigenmode for $S=0.143$, $\Reyn=360$,
$\DoTO=24.3$, together with $|\Real(\uh)|$ along $r=0.5$.
$(b)$ Spatial structure of the adjoint eigenmode associated with the direct
mode shown in $(a)$.
$(c)$ Structural sensitivity $\lambda$, as defined by~\eqref{eqn:structsens}.
$(d)$ Sensitivity $\bnabla_{\vect{\ub}} \, \omega_\text{i} = (\bnabla_{\ub} \, \omega_\text{i}, \bnabla_{\vb} \, \omega_\text{i})$ of the growth rate to modifications of the base flow.}
\label{fig:eigfunadjsens}
\end{figure}

\begin{figure}
\centering
\includegraphics[width=\textwidth]{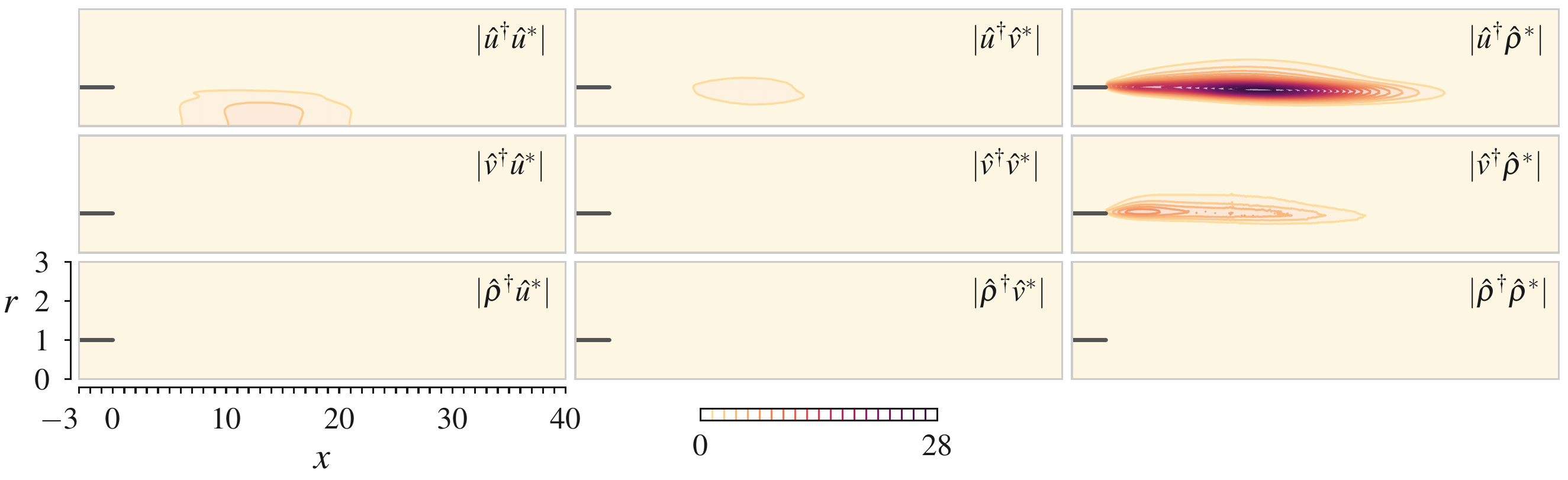}
\caption{%
Absolute value of the components of the structural sensitivity tensor $\tens{S}$
corresponding to the most unstable eigenmode for $S=0.143$, $\Reyn=360$, $\DoTO=24.3$.
}
\label{fig:senstens}
\end{figure}

The fact that the isolated mode is robust with respect to $x_\text{max}$
indicates that it must be quite distinct from the arc branch modes described
earlier. A first characterisation of this mode is attempted by inspecting its
spatial eigenfunction, shown in figure \ref{fig:eigfunadjsens}. Note that the results that
are shown were obtained with $x_\text{max}=60$, although for clarity only a fraction
of the domain (up to $x = 40$ and $r=3$) is shown. The top frame
$(a)$ represents a snapshot of the axial velocity perturbation. Quite in
contrast with the arc branch modes in figure \ref{fig:boxresonance}, the region
of significant eigenfunction amplitude is found to be well contained in the
centre of the domain. This is seen even clearer when looking at the inset where the
modulus of the perturbation values along a line $r=0.5$ is plotted in
logarithmic scale. At the inflow and at the outflow boundaries,
the perturbations are at least five orders of magnitude smaller than at their maximum location.

A numerical solution of the associated discrete adjoint eigenvalue problem
retrieves the complex conjugate counterparts of the direct eigenvalues, as
shown in figure~\ref{fig:Re360spectrum}, with high accuracy (arc branch and
isolated mode). The adjoint eigenfunction of the isolated mode is displayed in
figure \ref{fig:eigfunadjsens}$(b)$. It is strongly localised around the nozzle
edge, marking this region as being the most receptive to initial perturbations
for triggering the direct eigenmode.

Direct and adjoint eigenmodes may then be multiplied, according to \eqref{eqn:structsens}, in
order to estimate the flow region in which local feedback mechanisms contribute most to the
existence of the global eigenmode.  This quantity $\lambda$ is represented in
figure~\ref{fig:eigfunadjsens}$(c)$.  A well-localised maximum is found around $x=13$,
concentrated near the lower part of the shear layer; the potential core is also highlighted.
Comparing the individual components of the sensitivity tensor $\tens{S}$, shown in
figure~\ref{fig:senstens}, reveals that feedback proportional to the density perturbation
$\rho'$ into the axial and---to a lesser degree---radial momentum equations forms the strongest
contribution to changes in the eigenvalue.  Note that it does not tell us whether these changes
are stabilizing or destabilizing.  From inspection of the stability
equations~\eqref{eq:stabeq1}--\eqref{eq:stabeq3} at zero Richardson number, we can thus draw the
conclusion that the convection term $\vect{\ub}\bcdot\bnabla\vect{\ub}$ to which $\rho'$ is
proportional plays a highly important role for the growth rate and frequency of the isolated
eigenmode that is responsible for the self-sustained global oscillations in low-density jets.
The fact that the feedback has a stronger effect in the axial momentum equation than in the
radial momentum equation can easily be explained by remembering that the slenderness of this
moderately large Reynolds number jet flow implies that $\ub \gg \vb$ and consequently
$\ub\pfi{\ub}{x} \gg \ub\pfi{\vb}{x}$.

It is tempting, but hardly pertinent, to try to relate the spatial distribution of the
structural sensitivity to a supposed jet-column character of the eigenmode. The distinction
between jet-column and shear layer modes is meaningful in the context of a local analysis. A
physical examination of the active instability mechanisms in the isolated mode should ideally be
based on the role of the baroclinic torque, following the local analysis of
\citet{Lesshafft2007}.  However, it is not clear how the structural sensitivity could be
exploited for such a discussion.

It is noted from figure~\ref{fig:eigfunadjsens} that the shapes of the direct and adjoint
eigenmodes, as well as their pointwise product, compare well with the results of
\citet{Qadri2014}, shown in his figure~4.1. Recently, \citet{Qadri2015} have analysed
self-sustained oscillations in lifted diffusion flames. In their configuration, fuel with a
density 7 times smaller than that of the ambient is injected at a Reynolds number, based on the
present scales, of approximately 500, with a moderately steep velocity profile, $\DoTO=25$.
Given the similarities with the present set-up, it is not surprising that there are also many
similarities between their `mode A' and the isolated eigenmode under consideration here. For
example, the structural sensitivity component with the strongest contribution in their work is
that associated with changes in the mixture fraction feedback into the axial momentum equation
\citep[figure~6 of][]{Qadri2015}. Upstream of the diffusion flame, in the isothermal jet zone
where the sensitivity of their mode~A peaks, the mixture fraction is equivalent to the density,
so that their strongest sensitivity component is in fact the analogue of $\uh^\dag \rhoh^*$ in
the present analysis, which has indeed been shown to be the strongest contributor to the
structural sensitivity (figure~\ref{fig:senstens}). We would like to point out, however, that
the low-density jet region in \citet{Qadri2015} is bounded by a diffusion flame downstream, and
is thus not a canonical jet configuration.

Figure~\ref{fig:eigfunadjsens}$(d)$ shows the sensitivity of the growth rate of the global mode
to arbitrary modifications of the base flow. The magnitude $[(\bnabla_{\ub} \,
\omega_\text{i})^2 + (\bnabla_{\vb} \, \omega_\text{i})^2]^{1/2}$ is indicated by the colour,
whereas the arrows indicate the direction $(\bnabla_{\ub} \, \omega_\text{i}, \bnabla_{\vb} \,
\omega_\text{i})$ in which the base flow has to be modified to achieve a positive increment in
the growth rate, \ie to destabilize the global mode. The results are in line with those observed
by \citet{Tammisola2012} for unconfined wake flows, \ie a region of high sensitivity just
downstream of the injector, followed by on oscillatory pattern in a region that stretches from
approximately 5 to 20 radii downstream of the injector. This sensitivity measure can be further
separated in two parts: the sensitivity to changes in the base flow advection, and the
sensitivity to changes in the energy extraction from base flow gradients (`production'). It was
found (not shown in the figure) that the contribution of the production part corresponds to the
region of high sensitivity adjacent to the nozzle, while the advection part dominates in the
second region of high sensitivity farther downstream. A change in the velocity profiles just
downstream of the injector in the direction indicated by $(\bnabla_{\ub} \, \omega_\text{i},
\bnabla_{\vb} \, \omega_\text{i})$ would cause a thinning of the shear layer, increasing the
streamwise velocity gradient while bringing together the inflection points of the density and
velocity profiles. From a local stability point of view \citep{Lesshafft2010, Coenen2012}, it is
not surprising that such a change would cause a destabilisation of the flow.

\begin{figure}
\centering
\includegraphics[width=232pt]{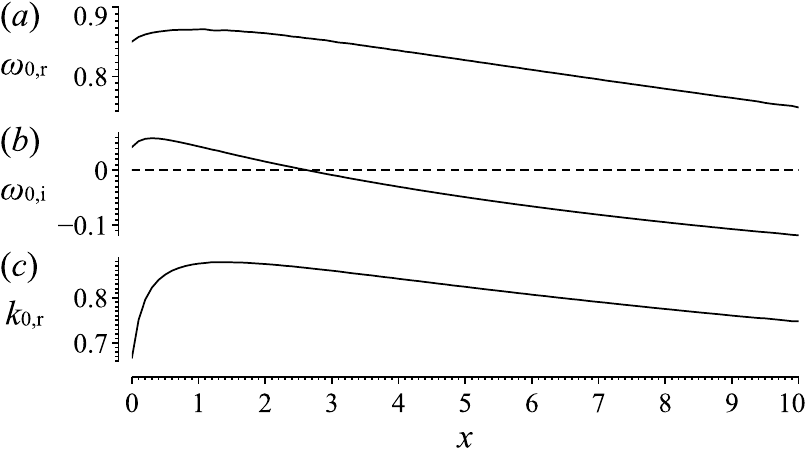}
\caption{%
Results of a local stability analysis for the case $S=0.143$, $\Reyn=360$,
$\DoTO=24.3$.}
\label{fig:local}
\end{figure}

If the isolated mode is not the result of non-local pressure feedback, which in
the present setting could only arise from spurious effects at the outflow
boundary, then it is expected to be linked to the presence of local absolute
instability. According to previous studies, for instance \cite{Lesshafft2007a},
global instability in jets requires an absolutely unstable region of
finite extent adjacent to the nozzle exit. To confirm the absolute character of
the local instability near the nozzle in the present base flow, a local
spatio-temporal stability analysis has been performed.
To that aim, at each downstream position $x$, the basic flow is assumed
to be locally parallel, with radial profiles of velocity
$\vect{\ub}(r) = (\ub(r), 0)$ and density $\rhob(r)$; small perturbations
are introduced as normal modes
$[\uh_l(r), \ui \vh_l(r), \ph_l(r), \rhoh_l(r)] \exp[\ui(k x - \omega t)]$, with complex
axial wavenumber $k = k_\ur + \ui k_\ui$ and complex angular frequency
$\omega = \omega_\ur + \ui \omega_\ui$. Here $k$, $\omega$, and $t$ are
non-dimensionalised using
$R^\ast$ and $U_m^\ast$.
Substitution of the normal modes into the equations of motion,
linearised around the steady base flow, yields a system of ordinary differential
equations that, together with
appropriate boundary conditions, provides a generalised eigenvalue problem
\citep[see, for instance,][]{Coenen2012},
to be interpreted as a dispersion relation $D(k,\omega; \Reyn, S, \DoTO, \ldots, \ub, \vb, \pb, \rhob) = 0$ between $k$ and $\omega$.
Here we are concerned with the absolute or convective character of the instability.
Therefore we need to find the spatio-temporal instability modes with zero group
velocity, \ie modes for which $\ud \omega / \ud k = 0$. The growth rate
$\omega_{0,\ui}$ of these is called the absolute growth rate and determines
whether the instability is convective, $\omega_{0,\ui}<0$, or absolute,
$\omega_{0,\ui}>0$. The condition $\ud \omega / \ud k = 0$ is equivalent to the
existence of a double root, or saddle point, in the complex $k$-plane,
$\left. \partial{D}/\partial{k} \right|_{k = k_0} = 0$. Among all the saddle points that
may exist, only the one with the largest value of $\omega_{0,\ui}$, while
satisfying the Briggs--Bers criterion, determines the large-time impulse
response of the flow
\citep[see, for instance,][and references therein]{Huerre2000}.
The numerical method used to determine $(\omega_{0,\ui},k_{0,\ui})$ is described in
\citet[Appendix B][]{Coenen2012}

Figure~\ref{fig:local} shows the streamwise variation of $\omega_{0,\ur}$,
$\omega_{0,\ui}$, and $k_{0,\ur}$.  Absolute instability
prevails over the interval $0\le x \le 2.6$. \citet{Couairon1999} and \citet{Lesshafft2006} showed
that when an absolutely unstable region is bounded by the jet outlet, the length $x_\text{AC}$ of this region needs to be sufficiently large for the global mode to be triggered.
\citet{Coenen2012} used the criterion $x_\text{AC} = C/\sqrt{\omega_\text{i}(x=0)}$ \citep{CHR88, Couairon1999} that contains a free parameter $C$. They found that $C=0.85$ gave good agreement with the experimental observations of \citet{Hallberg2006}. The same criterion would predict here that the length of the absolutely unstable region must be 4 radii. In figure~\ref{fig:local} we can observe that for this globally marginally stable flow, this length is approximately 3 radii.
From the spatially oscillating structure of the global mode of
figure~\ref{fig:eigfunadjsens}$(a)$, we can also estimate a wavenumber $k = 1.4$, to be compared
with the value $0.7 \simeq k_{0,\text{r}} \simeq 0.9$ in the absolutely unstable region of
figure~\ref{fig:local}$(c)$. The frequency of the global mode for this case is $\omega_r =
0.91$, whereas the local stability analysis results in a frequency that ranges from 0.85 to
0.87. The discrepancies between the two analyses can be attributed to the rapid spatial development of the flow (see figure~\ref{fig:baseflow}$(a)$) that violates the parallel flow hypothesis on which the local analysis is based. An clear example of this was found recently by \cite{Moreno2016} when studying buoyancy-driven flickering in diffusion flames.

\begin{figure}
\centering
\includegraphics[width=232pt]{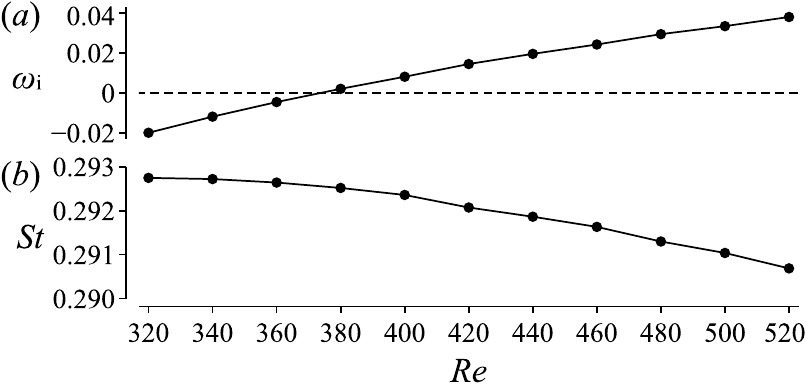}
\caption{%
Evolution of the isolated mode with the Reynolds number for a pure helium jet
($S=0.143$) and $\DoTO=24.3$.}
\label{fig:varRe}
\end{figure}

As the current value $S=0.143$ already represents the case of pure helium, in
figure~\ref{fig:varRe} the Reynolds number is varied in order to demonstrate
its influence on the isolated eigenvalue. Indeed, the growth rate $\omega_\ui$
is found to increase steadily with the Reynolds number, crossing into the
unstable half-plane before $Re=380$, while the Strouhal number remains almost
constant over the plotted interval of $Re$.

\begin{figure}
\centering
\includegraphics[scale=0.5]{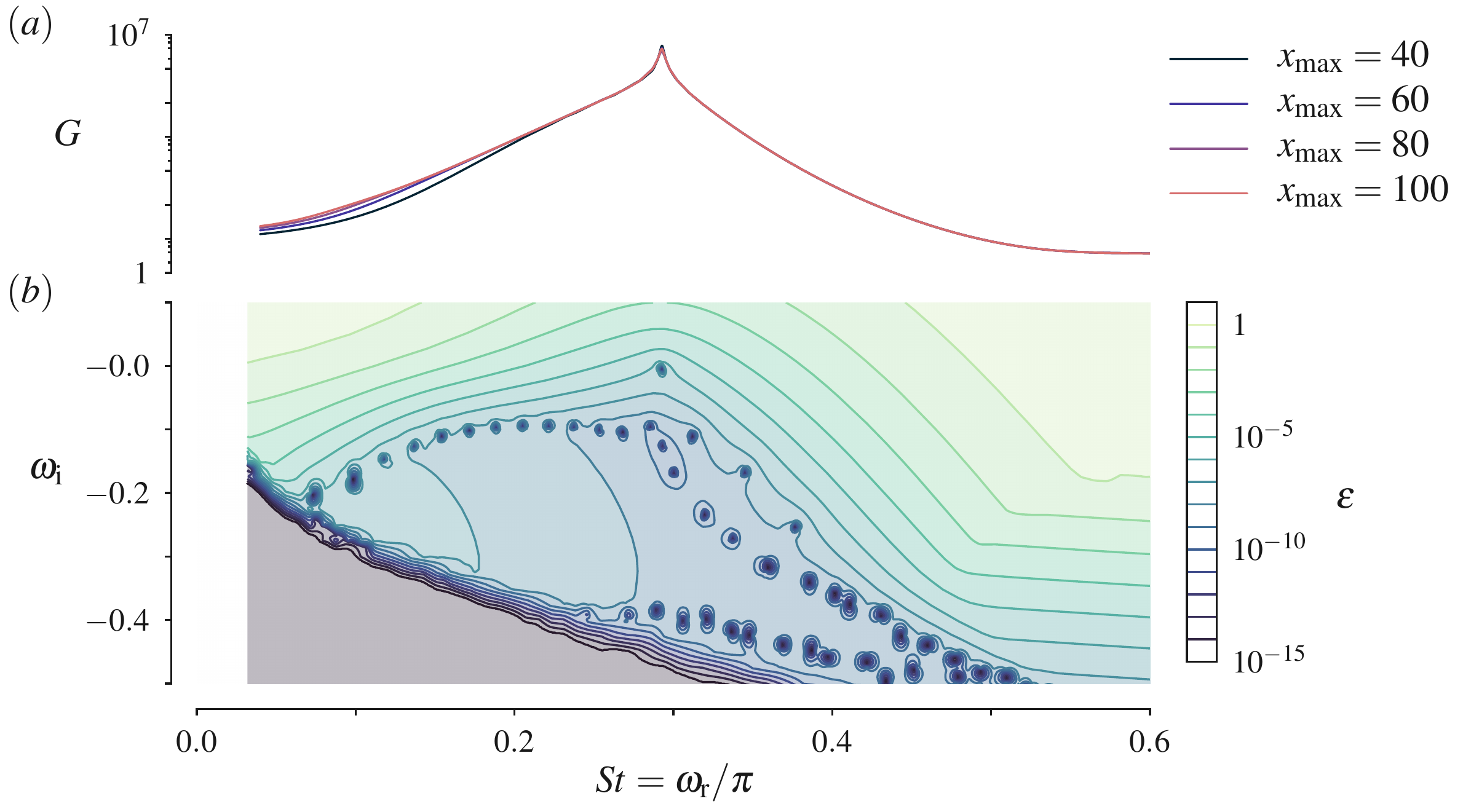}
\caption{%
Forcing response of a jet at $Re=360$, $S=0.143$ and $\DoTO=24.3$. $(a)$
Optimal gain as a function of real forcing frequency. Different colours denote
results obtained in numerical domains of different lengths, as indicated in the
legend. $(b)$ Pseudospectrum, computed as the inverse of the optimal gain for
complex forcing frequencies. The case $x_\text{max}=60$ is represented.}
\label{fig:Re360pseudo}
\end{figure}

The pseudospectrum of the $\Reyn=360$ helium jet is presented in figure
\ref{fig:Re360pseudo} for a complete comparison with the $\Reyn=1000$ case in
the previous section. Most features are shared by both configurations, in
particular the distinct variations of the energy gain in the regions above and
below the arc branch. The response to forcing at real frequency is well
converged in the case of the $\Reyn=360$ jet at all $x_\text{max}$ settings.
The most prominent difference with respect to the $\Reyn=1000$ case is a sharp
resonance peak in the energy gain near the frequency of the isolated mode. The
gain remains finite, because the isolated mode is still slightly stable. The
discussion of the arc branch provided in \S{}\ref{sec:Re1000} remains valid in
all aspects in the present configuration.

\subsection{Comparison with the experiment}
\label{sec:resultscompexp}

\begin{figure}
\centering
\includegraphics[width=\textwidth]{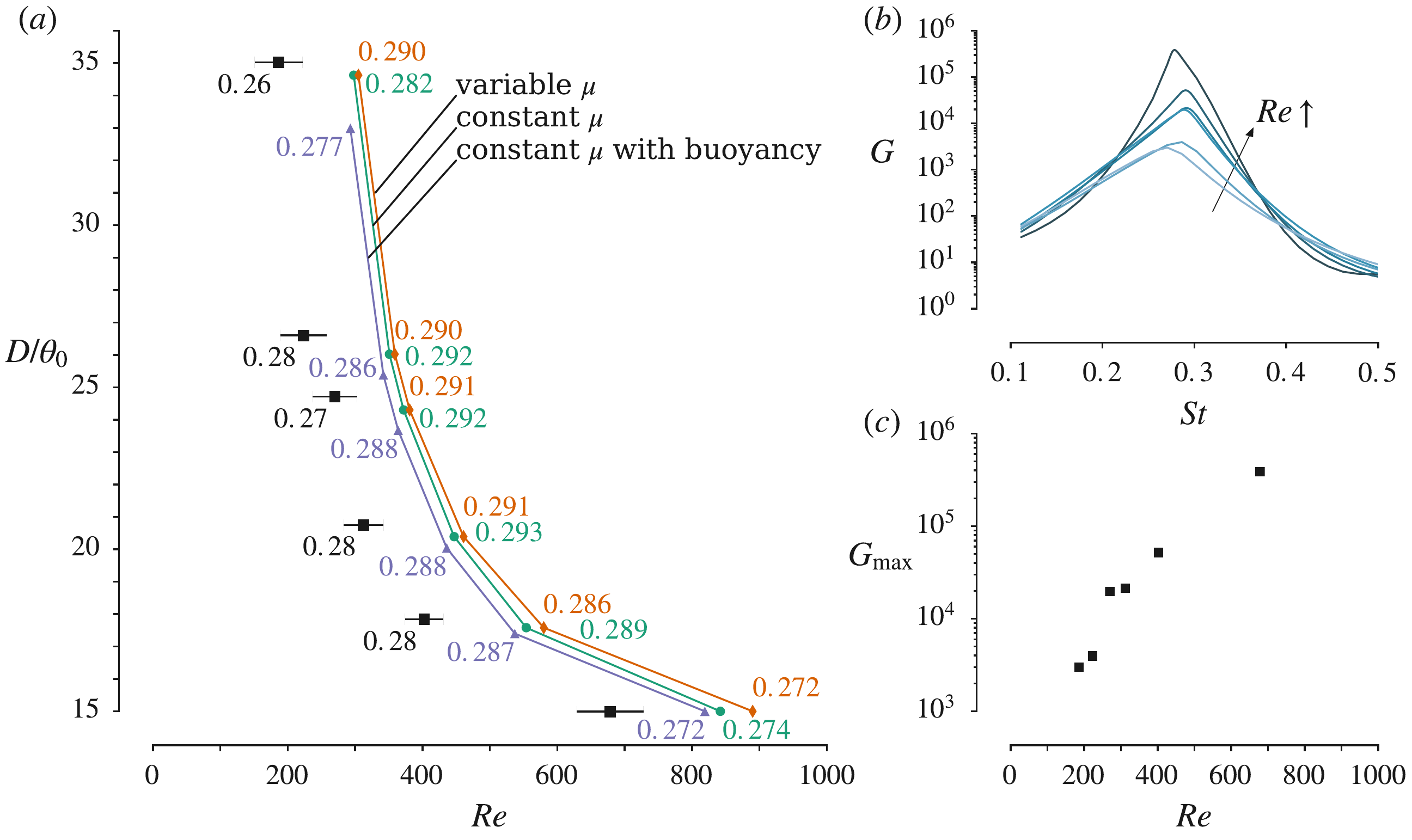}
\caption{%
$(a)$ Comparison with the experimental results of \citet{Hallberg2006} for
the onset of global instability in a pure $\He$ jet ($S=0.143$).
The experimental results are indicated with solid squares and
error bars, whereas the predictions based on the global mode analysis of the
present work are indicated by the solid circles. The addition of gravitational
effects in the analysis results in the transition points indicated by the
solid triangles, and taking into account the variation of viscosity with composition
results in the solid diamonds. The numbers next to the transition points
indicate the Strouhal numbers corresponding to the oscillating mode of the jet. 
$(b)$ Frequency response computed for conditions of the experimentally observed
transition points. $(c)$ The maximum optimal gain for the experimentally observed transition points.}
\label{fig:compexperiments}
\end{figure}

The pure helium jet at $\Reyn=360$ discussed in \S{}\ref{sec:Re360} has been
characterised as being nearly marginally stable. By tracking the values of the
control parameters $\Reyn$ and $\DoTO$ for which the growth rate of the
isolated mode $\omega_\ui$ is zero, a neutral curve can be constructed. In
figure~\ref{fig:compexperiments}($a$), such neutral curves in the $\Reyn-\DoTO$
plane are presented for pure helium jets ($S=0.143$). The solid circles
correspond to the results of the global mode analysis without taking into
account buoyancy effects. To the left of the transition points the jet is
globally stable ($\omega_\ui<0$) whereas to the right it is globally unstable
($\omega_\ui>0$).  These results are compared with the experimental
measurements of \citet{Hallberg2006}, indicated in
figure~\ref{fig:compexperiments}($a$) as squares with error bars. Care has been
taken to convert the experimentally obtained critical Reynolds numbers, based
on the centreline velocity, to Reynolds numbers based on the mean velocity.
Fair agreement is found between the experimental and the linear neutral curves,
but essentially the latter is shifted towards higher Reynolds numbers; in other
words, the onset of global instability in the linear calculations is delayed
with respect to the experimental observations.

Values next to the transition points in figure~\ref{fig:compexperiments}($a$)
indicate the Strouhal numbers near criticality. In the global mode
calculations, the Strouhal number $St$ is directly given by
the frequency $\omega_\ur$ of the marginally stable eigenmode, whereas in the
experiments, it was obtained by measuring the frequency of the self-sustained
oscillations under slightly supercritical conditions. It can be observed that
the agreement between the two is rather good, with relative differences smaller
than 10\%.

In an effort to explain the offset between the critical curve given by the
stability analysis and the experimental evidence, we assessed the influence of
buoyancy. Strictly speaking, while estimating the importance of the latter, the
characteristic length scale that should be used in the construction of the
Richardson number is not the radius $R^\ast$, but the development length of the
jet, of order $\Reyn R^\ast$. This modified Richardson number can be written
as $\Gras/\Reyn$. In the experiments of \cite{Hallberg2006} the Grashof number
for the pure helium jet ($S=0.143$) is $\Gras = 138$, and the marginal Reynolds
numbers lie in the range 200--700.  We can therefore expect that buoyancy has a
non-negligible effect in the experiments. Recomputing the critical curve with
the inclusion of the buoyancy term, with $\Gras = 138$, we obtained the solid
triangles of figure~\ref{fig:compexperiments}($a$). Indeed, adding buoyancy
destabilises the jet, and slightly improves the agreement with the experimental
data. Nevertheless, its influence is not sufficiently strong to explain the
offset between the stability analysis and the experiments.

A second physical aspect whose influence has been investigated is the variation of viscosity with composition. The viscosity of air is 11\% lower than that of helium, so that a small effect on the molecular transport can be expected, changing both the base flow and the stability properties. Figure~\ref{fig:compexperiments}($a$) shows that there is indeed a small shift of the transition curve (solid diamonds). The variable viscosity is seen to have a slightly stabilising influence. This seems counterintuitive, as a lower viscosity in the flow field due to variations with the composition would result in an higher local effective Reynolds number, and would therefore be expected to destabilize the flow. Nevertheless, subtle changes in the base flow profiles may just as well counteract this, eventually causing a net stabilization. A more detailed study may be interesting in flows where variations of the viscosity are stronger, such as heated jets or diffusion flames.

In figure~\ref{fig:compexperiments}($b$) we show the frequency response
computed at the transition points of \cite{Hallberg2006}. Because these points
are located to the left of the transition curve obtained with the global mode
stability analysis, the jet is globally stable under these conditions.
Nevertheless, the optimal gain is seen to be very high,
$\mathcal{O}(10^3-10^6)$, in a narrow band around the frequency associated
with the global mode (see also the discussion of figure~\ref{fig:Re360pseudo}).
This means that small perturbations that act as a
forcing to the jet may suffer very strong amplifications that may sustain a nonlinear
global oscillation of the jet at the frequency of maximum amplification. In an experiment
this could cause a shift in the observed critical value of the bifurcation parameter.
Applying this hypothesis to the present analysis would mean that, \emph{if} both the incoming noise and the necessary amplification threshold to sustain a nonlinear oscillation were constant over all experimental conditions, the computed frequency response at the experimental transition points should have values that are of the same order of magnitude. Nevertheless, figure~\ref{fig:compexperiments}($b$) shows that this is not the case here. In fact, the maximum gain is seen to be larger for the experimental conditions corresponding to higher Reynolds numbers (and lower $D/\theta_0$). It must be mentioned that these results correspond to the \emph{optimal} forcing, whose spatial structure varies from case to case, and might be very different from realistic noise present under experimental conditions. To rule out this variability the computations were repeated for a fixed spatial forcing distribution (a uniform distribution in the injection pipe), yielding similar results (not shown here) to the ones of figure~\ref{fig:compexperiments}($b$), but with generally lower values of the gain.

Finally, it is worth mentioning that, as the sensitivity to base flow modifications of figure~\ref{fig:eigfunadjsens}($d$) shows, small changes in the region just downstream of the injector
have a strong influence on the growth rate of the global instability. Although care has been
taken to mimic the experimental set-up of \citet{Hallberg2006}, some details, such as the exact nozzle shape and the sharpness of the nozzle lip, are hard to account for, and may as well play a role in the discrepancy between experiment and linear theory.


\section{Conclusions}
\label{sec:conclusions}

The present work gives, for the first time, a detailed account of the linear global
stability of low-density jets. By making use of the low Mach number approximation, all
dynamic effects of density variations in the limit of zero Mach number are retained, while
avoiding the numerically challenging necessity to resolve acoustic wave propagation.

We have found that when the spatial development of the jet flow is sufficiently fast
($\Reyn=360$ and $S=0.143$), an isolated eigenvalue dominates the global eigenvalue spectrum. This
mode has been shown to arise from the presence of absolute instability in light jets, documented
in numerous earlier studies \citep[e.g.][]{MonkewitzSohn,Lesshafft2007,Coenen2012}. It is this
mode that causes global instability in light jets at low Mach numbers. It has been found to
converge without much effort in our numerical calculations, in particular with respect to the
domain size. The structural sensitivity of this mode is concentrated in a confined region close
to the nozzle; according to \cite{GL07}, it is sufficient to resolve this flow region in order
to accurately compute the eigenvalue.

Unlike previous local stability analyses \citep[e.g.][]{Coenen2012}, in which the length of the
absolutely unstable region that is necessary to trigger a global mode introduces an unknown
parameter in the problem, the isolated global eigenmode is able the determine the critical
conditions for the onset of global instability in terms of the governing flow parameters without
any additional hypotheses. This has been employed to link the isolated mode to the
supercritical Hopf bifurcation that was observed in the helium jet experiments by
\cite{Hallberg2006}: the neutral curves for the onset of instability, in the experiments and in
the present linear analysis match, within reasonable accuracy, and the predicted Strouhal
numbers agree within ten per cent with the experimentally reported values at onset. The
bifurcation in the experiments takes place in situations that are characterised as slightly
subcritical in the linear framework. An additional destabilisation due to buoyancy effects has
been demonstrated to be insufficient in order to explain this offset.  Including the variation
of viscosity with composition has been shown to have a small stabilising effect, and is thus
also ruled out as an essential ingredient to explain the discrepancy.  An inspection of the
pseudospectrum however indicates that small perturbations may suffer a very strong amplification
in the slightly stable regime. According to linear theory, experimental low level noise might
therefore be amplified and observed as sustained coherent wavepackets. A dedicated detailed
study is required to assess this hypothesis, for example by comparing with direct numerical
simulations with a controlled low level forcing.


A second feature of the global eigenvalue spectra of low-density jets is a branch of
eigenvalues, called the arc branch, that, when the spatial development of the jet flow is
sufficiently slow ($\Reyn=1000$, $S=0.5$) may dominate the spectrum, hindering the detection of
the isolated mode. This arc branch is found to be highly sensitive to the numerical domain
size, consistent with numerous existing studies of jets \citep{Nichols2011,Garnaud2013PoF} and
boundary layers \citep{ehrenstein2005,ehrenstein2008, Akervik2008}. The present results suggest,
for the first time, that these eigenmodes are not only affected by domain truncation, but
that their very existence is dependent on spurious feedback from the outflow boundary. Two
observations support this conjecture: first, the arc branch aligns approximately with
isocontours of the pseudospectrum, suggesting a link between a given level of exogenous energy
input and the occurrence of arc branch modes; second, the associated eigenfunctions display an
integer number of wavelengths between inflow and outflow, suggesting a resonance condition. A
more detailed study to prove this conjecture is currently being carried out, but lies out of the
scope of the present work.



\begin{acknowledgments}

L.L. acknowledges financial support from ANR under the Cool Jazz project and from DGA under
contract LADX2331. W.C. and A.S. were supported by the Spanish MICINN under project
DPI2014-59292-C3-1-P and DPI2015-71901-REDT.

\end{acknowledgments}


\bibliographystyle{jfm}

\bibliography{references}

\end{document}